\documentclass[a4paper,12pt]{article}
 
\usepackage{graphicx}
\usepackage{cite}
\if@mathematic
   \def\vec#1{\ensuremath{\mathbf{#1}}}
\else
   \def\vec#1{\ensuremath{\mathchoice{\mbox{\boldmath$\displaystyle#1$}}
                              {\mbox{\boldmath$\textstyle#1$}}
                              {\mbox{\boldmath$\scriptstyle#1$}}
                              {\mbox{\boldmath$\scriptscriptstyle#1$}}}}
\fi
\newcommand{\rmi}{\textrm{i}}
\newcommand{\rmd}{\textrm{d}}
\newcommand{\be}{\begin{equation}}
\newcommand{\ee}{\end{equation}}
\newcommand{\beq}{\begin{equation}}
\newcommand{\eeq}{\end{equation}}
\newcommand{\slh}{\!\!\!/} %Dirac slash
\newcommand{\slhh}{\!\!\!\!/} %Dirac slash (for large letters)
\newcommand{\slhhh}{\!\!\!\!\!/} %Dirac slash (even larger)

\begin{document}
 
\thispagestyle{empty}   % to suppress the page number on the first page

%\hfill\fbox{\begin{minipage}{5cm}
% $ $RCSfile: emufizika.tex,v $ $ \\
% $ $Date: 2001/04/19 15:46:15 $ $ \\
% $ $Revision: 1.14 $ $
%\end{minipage}}

\begin{flushleft}
\small
\today \\
OSLO-TP-5-01\\
ZAGREB-ZTF-01/02
\end{flushleft}

\begin{center} \Huge \bf
The double radiative annihilation of the heavy-light 
fermion bound states\footnote{Dedicated to Prof. Ksenofont Ilakovac on the occasion of his seventieth birthday}
 \\[1cm]
\large
J. O. Eeg\footnote{j.o.eeg@fys.uio.no}$^{\ast}$,
K. Kumeri\v{c}ki\footnote{kkumer@phy.hr}$^{\dagger}$,
and  I. Picek\footnote{picek@phy.hr}$^{\dagger}$
\setcounter{footnote}{0}
\end{center}
\vspace*{5mm}
\noindent
$^{\ast}$ {\small\rm 
Department of Physics, University of Oslo, N-0316 Oslo, Norway} \\
$^{\dagger}$ {\small\rm Department
 of Physics, Faculty of Science, University of Zagreb, POB 331,
 Bijeni\v{c}ka cesta 32, HR-10002 Zagreb, Croatia}\\

\begin{abstract}
We consider the double-radiative decays of heavy-light
QED and QCD atoms, $\mu^{+} e^{-} \rightarrow \gamma\gamma$ and
$\bar{B}^{0}_{s} \rightarrow\gamma\gamma$. Especially, we take under
scrutiny contributions coming from
operators that vanish on the free-quark mass shell.  
We show that by field redefinitions these operators are converted
into contact terms attached to the bound state dynamics. 
A net off-shell
contribution is suppressed with respect to the effect of the well known
flavour-changing magnetic-moment operator by the bound-state binding
factor. The negligible off-shellness of the weakly bound
QED atoms becomes more relevant for strongly bound QCD atoms. We analyze
this off-shellness in model-approaches to QCD, one of them
enabling us to keep close contact to the related effect in QED.
We also comment on the off-shell effect in the corresponding process
$\bar{B}_d \to K^* \gamma$, and discuss possible hindering of the claimed
beyond-standard-model discovery in this decay mode.
\end{abstract}



\newpage

\section{Uses of a comparative study of the off-shell effects in
  QED and QCD atoms}

  Off-shell effects are known to be quite elusive. The most famous
measured effect, dubbed Lamb shift \cite{LaR47,BeS57}, appears in atomic 
physics. It is
represented by the atomic level shift on account of a tiny difference
in the self-energies of the free electron and of the
electron bound in the H-atom. 
Half a century after its discovery, investigations of Lamb shift still
provide a precision test of bound-state QED \cite{Ka00}.
Now, an experimental uncertainty of 3 ppm in laser experiments is
essentially smaller then the 10 ppm theoretical inaccuracy due to poor
knowledge of the proton charge radius. In this situation the study
of unstable leptonic atoms becomes competitive to the study of
hydrogen.

 In a study of the inter-nucleon potential there have been some early
expectations \cite{AsM49} to reveal the off-shell parts of the nucleon-nucleon
bremsstrahlung amplitude. Fearing and Scherer
\cite{FeS00,Fe00} excluded this possibility by subsuming such off-shell
amplitudes into redundant terms \cite{We95} that can be rotated by field
redefinitions into so called contact terms.
The parity violating anapole terms \cite{MuH91} might be one exception, 
deserving a separate study.

However, particle physics provides new microscopic in\-te\-ract\-i\-ons 
leading to
potentially interesting new contact terms. The most famous one is the
anomalous $\pi^0 \gamma \gamma$ coupling. As explained in some
detail in \cite{EeP94c} this coupling can be viewed as an off-shell effect. 
On top  of the QCD binding of the
quark-antiquark atom, the triangle quantum loop dominated by far off-shell
quarks produces an anomalous coupling responsible for the $\pi^0$
decay. Such manifestation of the off-shellness motivates us to study
the two-photon annihilation of atoms in general.

  The simplest ``total disintegration'' of an ``atom'' occurs when it 
consists of a
particle-antiparticle pair, like in the case of the true QED-atom,
positronium\footnote{A revival of positronium \cite{Ri81}
appeared after the discovery of the QCD atoms, together with the
recognition that the first proposal of the positronium (termed
``electrum'') has been given as early as in 1934 \cite{Mo34},
immediately after the discovery of the positron by Anderson.}.  Actually,
a comparative study of  QED and QCD atoms has been very fruitful in
the early days of quarkonia. 
A total disintegration of an atom consisting of different fermions is
more subtle. It happens on account of the flavour changing (FC) processes
familiar from weak interactions. 
While the transitions among charged quarks are well known, the lepton-flavour
violating (LFV) transitions among charged leptons is an open urgent issue,
stimulated by accumulated indication of the neutrino oscillations. In order
to benefit from the crossfertilization of different fields we pursue here
the comparative
study of annihilation in the \emph{heavy-light} QED and QCD systems.

Such a comparative study
throws new light on the off-shell nonperturbative effects of
valence quarks, studied first by two of us in the case of the double
decays of the $K_L$ \cite{EeP94,EeP93} and $\bar{B}_{s}$ meson \cite{EeP94b}.
Subsequently, this study has been continued within the specific bound state
models, both for $K_L \rightarrow 2\gamma$ \cite{KeKKP95} and for 
$\bar{B}^{0}_{s}  \rightarrow 2\gamma$ \cite{VaE98}.
In these papers it was explicitly demonstrated that operators that
vanish by using the perturbative equations of motion gave nonzero 
contributions for processes involving bound quarks. 
The purpose of the present paper is to elaborate to more detail our
more recent study \cite{EeKP00} which accounts for  similar effects
for the bound leptons.

\subsection{The relativistic QED atom}

 Our starting point is the Lagrangian density for the two fermions
in the absence of the FC transitions. Thus the light particle
(electron $e^-$ of mass $m$) and a heavy positively charged particle
(say muon $\mu^+$ of mass $M$) interact only through electromagnetic field, 
as given by the last term in
\begin{equation}
{\cal L}={\cal L}_{e}+{\cal L}_{\mu} -
 \frac{1}{4}F^{\alpha\beta}F_{\alpha\beta} - J^{\alpha}A_{\alpha} \;.
\end{equation}
The Dirac Lagrangian
\begin{equation}
{\cal L}_i = \bar{\psi}_i \left[ \frac{\rmi}{2}\gamma^{\alpha}\frac{
\stackrel{\leftrightarrow}{\partial}}{\partial x^{\alpha}} - m_i \right]
\psi_{i} \;\; ; \quad \stackrel{\leftrightarrow}{\partial} =
 \stackrel{\rightarrow}{\partial} -
\stackrel{\leftarrow}{\partial} \;\;  ,
\end{equation}
for a given particle ($i=e, \mu$) leads to the Dirac equations for 
$\psi_i$ and $\bar{\psi}_i$ treated as independent fields
\begin{equation}
  \bar{\psi}_i(\rmi\stackrel{\leftarrow}{\partial} \!\!\!\!\! / + m_i) =0
\;, \quad (\rmi \partial \!\!\! /  -m_i)\psi_i = 0 \;.
\end{equation}
Imposing the Coulomb (radiation) gauge, $\nabla\cdot\vec{A}=0$, one is able to
solve for $A^{0}$ (eliminate it from the Lagrangian), leading to
\begin{equation}
{\cal L}={\cal L}_{e}+{\cal L}_{\mu} +\frac{1}{2}(\vec{E}^{2}_{\perp}
- \vec{B}^2) + \vec{J}\cdot\vec{A} - \frac{1}{2} \int 
\frac{\rmd^3 r'}{4\pi} \frac{\rho(\vec{r},t)\rho(\vec{r}',t)}{
|\vec{r}-\vec{r}'|} \;.
\end{equation}
Here the last two terms can be expressed in terms of $\rho$ and $\vec{J}$
components of the fermion current
\begin{equation}
J^{\alpha}=e(\bar{\psi}_{\mu} \gamma^{\alpha} \psi_{\mu} -
          \bar{\psi}_e \gamma^{\alpha} \psi_e) \;.
\end{equation}
The corresponding Hamiltonian, after neglecting the self-energy terms in
the Co\-u\-lomb interaction, has the form \cite{Gr93}
\begin{equation}
H(x)=H(x)_{\rm Atom} + H(x)_{\rm Rad} +
     H(x)_{\rm Coulomb-inst} + H(x)_{\rm int} \;.
\end{equation}
where  
\begin{equation}
H_{\rm Atom}=H_{\mu} + H_e  \;,
\end{equation}
contains the relativistic  fermion contributions
\begin{equation}
H_{e(\mu)}=\int \rmd^3 r\, {\cal H}^{0}_{e(\mu)} = \int \rmd^3 r\,
\psi^{\dagger}_{e(\mu)}(x)
  \big[ -\rmi \vec{\alpha}\cdot\nabla + m_{e(\mu)}\beta \big]
  \psi_{e(\mu)}(x) \;.
\end{equation}
The electromagnetic piece splits into the radiation part
\begin{equation}
H_{\rm Rad}=\frac{1}{2}\int \rmd^3 r 
 \big[\vec{E}^{2}_{\perp}(x)+\vec{B}^2(x) \big] 
\end{equation}
containing the relevant electric and magnetic fields
\begin{equation}
  \vec{E}_{\perp}=-\frac{\partial\vec{A}}{\partial t} \;, \quad
   \vec{B}=\nabla \times \vec{A}\;,
\end{equation}
and the instantaneous Coulomb term
\begin{equation}
H_{\rm Coulomb-inst}=\frac{1}{4\pi}
\int \frac{\rmd^3 r \, d^3 r'}{|\vec{r}-\vec{r}'|}
J^{0}_{\mu}(\vec{r},t) J^{0}_{e}(\vec{r}',t) \;.
\end{equation}
The relativistic QED atoms can be
treated to all orders by solving exactly the Dirac equation with a
Coulomb interaction. This means solving the Dirac equation with 
$V(x)=\gamma_{0}V_{c}(x)$ ($V_c$ denotes the Coulomb potential)
\begin{equation}
  \big[ \rmi \partial\slh + V(x) - m_i \big] \psi_{i} = 0 \;.
\label{CoulDir}
\end{equation}
Correspondingly, the fermion propagator in external field  reads
\begin{equation}
  \big[ \rmi \partial\slh + V(x) - m_e \big] S^{e}_{F}(x, y)=
 \delta^{(4)}(x-y) \;.
\label{lightprop}
\end{equation}
Thus, in contrast to the free-particle propagator, 
the propagator for the bound fermion,
\begin{equation}
\rmi S_{F}(x, y)= \theta(x_0 - y_0)\sum_{n,\sigma}
\psi^{(+)}_{n,\sigma}(x)\bar{\psi}^{(+)}_{n,\sigma}(y)
- \theta(y_0 - x_0)\sum_{n,\sigma}
\psi^{(-)}_{n,\sigma}(x)\bar{\psi}^{(-)}_{n,\sigma}(y) \; ,
\end{equation}
should require the sum over all possible excited states, which appear
when decomposing the fermion field in terms of a complete set of positive and
negative energy eigenfunctions:
\begin{equation}
  \psi_e(x)=\sum_{n,\sigma} \left\{
  b_{n,\sigma}\psi^{(+)}_{n,\sigma}(x) +
  d^{\dagger}_{n,\sigma}\psi^{(-)}_{-n,-\sigma}(x) \right\} \;.
\end{equation}
The solutions of the free Hamiltonian
\begin{equation}
H_{0}=H_{\rm Atom} + H_{\rm Rad} + H_{\rm Coulomb-inst} 
\end{equation}
form a complete set of stationary states $|a, N\rangle$,
expressed as a direct product of atomic wavefunctions
$\psi_{a}$ and the photon Fock states
\begin{equation}
  |a, N\rangle = \psi_{a}(r)|N\rangle \;.
\end{equation}
When the interaction is turned on, one should make a replacement
\begin{equation}
   H_{0} \rightarrow H=H_{0}+H_{I}(t),
\end{equation}
and the pertinent states cease to remain stationary. Their evolution in time
in practice means that the excited states decay under the influence
of the QED interaction
\begin{equation}
  H_{I}(t)=-\int \rmd^3 r \Big[ \vec{J}_{p}(\vec{r},t)+
 \vec{J}_{e}(\vec{r},t) \Big] \cdot \vec{A}(\vec{r},t)
\label{eq:QEDint}
\end{equation}
into other states  $|b, N\rangle$, where $N$ photons are emitted.

In addition to the ordinary interaction (\ref{eq:QEDint}),
in the next subsection we shall consider
possible additional interactions (\ref{eq:lagg}), involving
the flavour changing $\mu\leftrightarrow e$ transition. This will
enable the atom to disintegrate completely. The lowest order
disintegration requires $\mu$ and $e$ overlap, happening when $l=0$,
i.e. a decay from S states. The decay from $l>0$ corresponds to a
cascade down to S state, followed by the decay from there --- a higher
order process which we do not need to consider in what follows.

\subsection{Some motivation for scrutinizing muonium}

There has been a considerable revival of the interest in muonium 
($\textrm{Mu} =\mu^+e^-$ system) in view of the very precise
measurements in this system. At the same time, the theoretical
predictions are plagued by the nonperturbative bound-state effects.
The only known way to achieve the required precision for the
bound states is by expanding around a nonrelativistic limit. Such
methods, like Non-Relativistic Quantum Electrodynamics \cite{CaL86,Le97}
start from the bound state described by a Schr\"{o}dinger 
wave function, and build up corrections in terms of the relative
velocity of the components.

For the muonium at hand our analysis and results bear a close
analogy to the correction to the muon lifetime due to muonium
formation, reported in \cite{CzLM00}.
In this system electron and muon have r.m.s. velocities
$\beta_{e}=\alpha\simeq 1/137$, and $\beta_{\mu}\simeq \alpha m_e/m_\mu
\simeq 3.5\cdot 10^{-5}$. In terms of these parameters the
bound-state corrections acquire a form $\alpha^n (m_e/m_\mu )^m$, where
the corrections up to $n+m=4$ matter in practice.
The current world average for the muon lifetime measurements 
\cite{PDG00}
\be\label{eq:mu_to_e}
\tau_\mu
= \, 2.19703(4)\times 10^{-6}\;\mbox{\rm s}  \, ,
\nonumber
\ee
has an uncertainty of only 18 ppm. 
In order to benefit from an improvement the measurement of $\tau_\mu$ (and
thereby of $G_\mu$)  by a factor of  
20 (i.e. reducing its uncertainty to only $\pm 1$ ppm)
a knowledge                                                        
of modification of  $\tau_\mu$  due to  the formation of muonium is
required.

The reexamination 
of the muonium bound state effect \cite{CzLM00} showed only a tiny effect, the
$\simeq 6\cdot 10^{-10}$ correction to the lifetime
\begin{center}
$\, \tau_{\textrm{\scriptsize Mu}} = \tau_\mu  \left(1 + {\alpha^2\over 2} {m_e^2\over m_\mu^2} 
\right).$
\end{center}
This negligible overall shift is in contrast to 
relatively large ${\cal O}(\alpha  {m_e\over m_\mu})$    
velocity effects on the spectrum \cite{CzLM00}. 
In the present work we are pointing out the off-shell effects (\ref{totalG}) 
in the 
radiative annihilation of muonium which are in between of these two.

For completeness, let us mention that besides the mentioned radiative
annihilation, there is also the $W$-exchange annihilation
Mu$\to \nu_e \bar{\nu}_\mu$ (the analog of $\mu^- p$ capture) with
the rate
\begin{equation}
 \Gamma(\textrm{Mu}\to\nu_e \bar{\nu}_\mu ) = 48 \pi \left(
\frac{\alpha m_e}{m_\mu}\right)^3 \Gamma(\mu^+ \to e^+ \nu_e \bar{\nu}_\mu) \;.
\end{equation}
Still, it leads to a miniscule branching ratio $\approx 7\cdot 10^{-12}$,
and moreover is restricted to the orthomuonium decay, which is out of
our scope here.
Let us note that some interest in radiative orthomuonium decay might
come from the three-photon analog decays: the puzzling discrepancy in
orthopositronium (a brief sketch of the recent status can be found in
\cite{AdFS00}) and the surprising suppression in theoretical
estimate for $K_L\to 3\gamma$ \cite{HeMS94}. These three-photon
decays provide three-party entanglement similar to the one in
quantum optics \cite{BoPDWZ99}.

Of course, the muonium annihilation involves the LFV
transition which is a matter of the beyond the standard
model (BSM) physics. Such lepton-number violating interaction induces
simultaneously a $\mu \to e \gamma$ transition \cite{ChL84}, so that the
unknown details will cancel in the ratio of these two processes.

For the heavy-light muonium system $\mu^+ e^-$ 
(where $m_{\mu}\equiv M \gg m_e \equiv m$), the bound-state calculation 
corresponds to that of the relativistic hydrogen. Thereby we distinguish
between the Coulomb field responsible for the binding, and the radiation field
\cite{Sa67}
participating in the flavour-changing transition at the pertinent 
high-energy scale.
In this way, the radiative disintegration of an atom becomes tractable 
 by implementing the two-step treatment \cite{ItZ80}:
``neglecting at first annihilation to compute the binding and then neglecting
binding to compute annihilation''.
This factorization of scales was introduced for the first time by
Wheeler \cite{Wh46}.
For the muonium atom at hand, the binding problem     
is analogous to a solved problem of the H-atom. 
In this way we  avoid the 
relativistic bound state problem, which is a difficult
subject, and we have no intention to contribute to it here.

The mentioned two-step method is known to work well for 
 disintegration (annihilation) of the simplest
QED atom, positronium.
Generalization of this procedure to muonium means  that the two-photon
decay width of muonium is obtained by using
\begin{equation}
\Gamma=\frac{|\psi(0)|^2\: |{\cal M}(\mu^{+}e^{-}\to\gamma\gamma)|^2}
            {64 \pi M m} \;,
\label{width}
\end{equation}
where $|\psi(0)|^2$ is the square of the 
bound-state wave function at the origin. 
After this factorization has been performed, the rest of the problem
reduces to the evaluation of the scattering-annihilation invariant
amplitude ${\cal M}$.
In the case of positronium this expression will involve equal
masses ($M$=$m$), and 
  the invariant
amplitude which for a positronium annihilation at rest has the textbook
form~\cite{Na90}
\begin{equation}
{\cal M}=\frac{\rmi e^2}{2m^2}\bar{v}_{s}(p_2)\Big\{
\epsilon\slh_{2}^{*}\epsilon\slh_{1}^{*}k\slh_{1}+
\epsilon\slh_{1}^{*}\epsilon\slh_{2}^{*}k\slh_{2}
\Big\}
   u_r(p_1) \;.
\label{positronium}
\end{equation}
Only the antisymmetric piece in the decomposition of the product of
three gamma matrices above
\begin{equation}
\Big\{ \;\; \Big\} \to \rmi
 \epsilon ^{\mu \nu \alpha \beta} \gamma_5 \gamma_{\beta}
(k_1 - k_2)_{\alpha}
(\epsilon ^*_1)_{\mu}(\epsilon ^*_2)_{\nu} \;,
\label{parapositronium}
\end{equation}
contributes to the spin singlet parapositronium two-pho\-ton 
annihilation. This selects
($\vec{\epsilon}^{\ast}_{1}\times\vec{\epsilon}^{\ast}_{2}$),
a CP-odd configuration 
of the final two-photon state.
We will see that for muonium annihilation also the CP-even
$\vec{\epsilon}^{\ast}_{1}\cdot\vec{\epsilon}^{\ast}_{2}$
configuration contributes.

The paper is organized as follows: In section 2 we consider the
quantum field treatment of the annihilation process
$\mu^{+}e^{-} \to \gamma \gamma$  in 
 arbitrary external field(s).
In section 3 we relate the binding forces to the external fields of section 2.
In section 4 we perform the calculation of $\bar{B}_s\to\gamma\gamma$ in several
different QCD models. In addition we consider the related off-shell
bound-state effects in $\bar{B}_d\to K^* \gamma$ decay.
In section 5 we present our conclusions.

\section{Flavour-changing operators for $\mu^{+}e^{-} \to \gamma \gamma$}

\begin{figure}
\centerline{\includegraphics[scale=0.8]{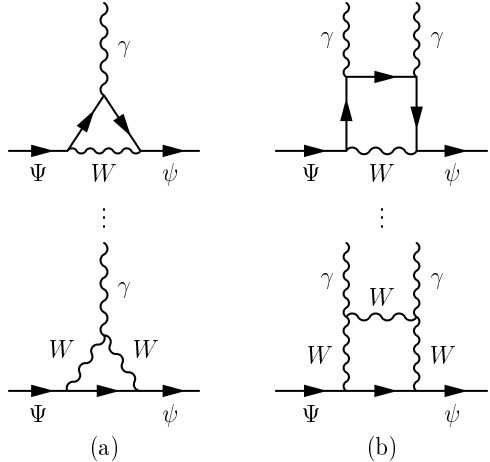}}
\caption{The examples of the one-particle-irreducible diagrams leading to the
 double-radiative flavour-changing transitions. Only the second-row diagrams
exist for the leptonic case.
\label{1PI}}
\end{figure}

Augmenting the electroweak theory by LFV enables the one- and two-photon
radiative decays $\mu\to e\gamma$ and $\mu\to e\gamma\gamma$.
Accordingly, the double-radiative 
transition is triggered by two classes of one-particle irreducible
diagrams (Figs. \ref{1PI}a and b), 
related by the Ward identities. 
After integrating out the heavy  particles in the loops, 
these one-loop electroweak  transitions can be combined  into
an effective Lagrangian \cite{EeP93},
\beq
{\cal{L}}(e \rightarrow \mu)_{\gamma} \, = \,  B \, 
\epsilon^{\mu \nu \lambda \rho}
F_{\mu \nu} \, ( \bar{\Psi} \; \rmi \stackrel{\leftrightarrow}{D_{\lambda}}
 \gamma_{\rho} L\, \psi )  \, + \mbox{h.c.}  \; ,
\label{eq:lagg}
\eeq
where muon and electron
 are described by quantum fields $\Psi = \psi_{\mu}$ and $\psi= \psi_e$.
Correspondingly, for $\bar{B}^{0}_{s} \rightarrow 2 \gamma$, the involved fields
are $\psi_s=s$ and $\psi_b=b$.

In our case, we do not need to specify the physics behind the 
lepton-flavour-violating  transition in (\ref{eq:lagg}). 
For instance, the strength $B$ might contain  
Maki-Nakagawa-Sakata \cite{MaNS62} parameters, 
analogous to the Cabibbo-Kobayashi-Mas\-ka\-wa parameters $\lambda_{\rm CKM}$
in the quark sector.

Keeping in mind that the fermions in the bound states are not
on-shell, we are not simplifying the result of the electroweak
loop calculation by using the perturbative eq\-uation of motion.
Thus the effective Lagrangian (\ref{eq:lagg})
obtained within perturbation theory 
splits into the on-shell magnetic transition operator ${\cal{L}}_{\sigma}$
\beq
{\cal{L}}_{\sigma} (1 \gamma) = B_{\sigma} \bar{\Psi} 
 \, (M  \sigma \cdot F L + 
 m\,  \sigma\cdot F R) \,  \psi \, + \mbox{h.c.} \; , 
\label{eq:lagmag}
\eeq
and an off-shell piece ${\cal{L}}_F$ \cite{EeP93} 
\beq
{\cal{L}}_{F} = B_F\, \bar{\Psi} [(\rmi\stackrel{\leftarrow}{D}\slhhh - M) \,
 \sigma \cdot F L + \sigma \cdot F R (\rmi\,
D\slhh -m)] \psi \, + \mbox{h.c.} \; , 
\label{eq:laggg}
\eeq
where $\sigma\cdot F$ denotes $\sigma_{\mu \nu} F^{\mu \nu}$, and
$L=(1-\gamma_5)/2$ and $R=(1+\gamma_5)/2$ denote left-hand and right-hand
projectors.
To lowest order in QED (or QCD) $B_F = B_\sigma = B$, but in general they are
different due to different anomalous dimensions of the operators in
(\ref{eq:lagmag}) and (\ref{eq:laggg}). Let us note that the 
off-shell part ${\cal{L}}_{F}$ has zero anomalous dimension \cite{EeP94b}.

\begin{figure}
\centerline{\includegraphics[scale=0.8]{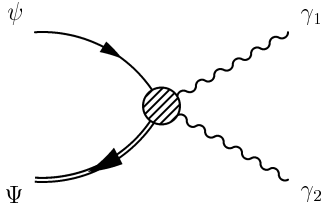}}
\caption{The two-photon contact (seagull) diagram that can be rotated
  away by a field redefinition.\label{seagull}}
\end{figure}

By decomposing the covariant derivative, 
$\rmi D\slhh =  \rmi \partial\slh - e A\slh$, in the off-shell operator
(\ref{eq:laggg}), we separate the one-photon piece
\beq
{\cal{L}}_F(1 \gamma)
 = B_F \, \bar{\Psi} 
[(\rmi \stackrel{\leftarrow}{\partial\slh} - M) \,
 \sigma \cdot F L + \sigma \cdot F R (\rmi
\partial\slh -m)] \psi \, + \mbox{h.c.}\; ,
\label{eq:lag1g}
\eeq
from the two-photon piece
\beq
{\cal{L}}_F(2  \gamma)
= B_F\, \bar{\Psi} [- e A\slh \sigma \cdot F L +
\sigma \cdot F R (- e A\slh)]\psi \, + \mbox{h.c.}\; .
\label{eq:lag2g}
\eeq
The amplitude for the two-photon diagram (Fig. \ref{seagull}) is given by
\beq
A_a = \rmi \int \rmd^4x \, {\cal{L}}_F(2  \gamma)
 = A^L_a + A^R_a \;,
\label{eq:lag2gD}
\eeq
in an obvious notation. 
The single-photon off-shell Lagrangian ${\cal{L}}_F(1 \gamma)$
leads to the amplitude with the heavy particle in the propagator
\begin{eqnarray}
A_b &=&  \rmi \,  B_F \, \int \!\! \int \rmd^4x \, \rmd^4y \, \bar{\Psi}(y) 
\Big[-\rmi e  A\slh_{2}(y)\Big]\, \rmi S_F^{(\mu)}(y,x) \nonumber \\
&&\hspace*{-2em}\times\bigg[(\rmi \stackrel{\leftarrow}{\partial\slh_x} - M) \,
 \sigma \cdot  F_1(x) L   + \sigma \cdot  F_1(x) R 
  (\rmi \partial\slh_x -m)\bigg] \psi(x) \; ,
\label{eq:lag1gDb}
\end{eqnarray}
and a similar amplitude with the light particle in the propagator
\begin{eqnarray}
A_c & = &  \rmi \,  B_F \, \int \!\! \int \rmd^4x \,\rmd^4y \, \bar{\Psi}(x) 
 \bigg[(\rmi \stackrel{\leftarrow}{\partial\slh_x} - M) \,
 \sigma \cdot  F_1(x) L +  \nonumber \\
&& \hspace*{-2em} \sigma \cdot  F_1(x) R (\rmi
\partial\slh_x -m)\bigg]  
\times \, \rmi S_F^{(e)}(x,y) \Big[-\rmi e A\slh_2(y)\Big] 
\psi(y) \; .
\label{eq:lag1gDc}
\end{eqnarray}
The subscripts 1 and 2 distinguish between the two photons. It is
understood that a term with the $1 \leftrightarrow 2$ subscript
interchange should be added in order to make our result symmetric in the
two photons.

Within the quantum field formalism, the sum of the equations
(\ref{eq:lag2gD}), (\ref{eq:lag1gDb}) and (\ref{eq:lag1gDc})
describes the process
$\mu^{+}e^{-} \to \gamma \gamma$, or 
$\mu  \to  e \gamma \gamma$.

Let us now be very general, and assume that both particles
($e$ and $\mu$) feel some kind of external field(s)  represented by
$V_{(e)}$ and $V_{(\mu)}$, and obey 
 one-body Dirac equations
\begin{equation}                                                               
  \big[ \rmi \partial\slh - V_{(i)}(x) - m_{(i)} \big] \psi_{(i)} = 0 \;,
\label{CoulDir2}                                                                
\end{equation}
for $i= e$ or  $\mu$ (in general $V_{(i)}= \gamma_{\alpha} \, V^{\alpha}_{(i)}$)
, and accordingly
the particle pro\-pa\-ga\-tors
$S_F^{(i)}$ satisfy:
\begin{equation}                                                               
  \big[ \rmi \partial\slh - V_{(i)}(x) - m_i \big] S^{(i)}_{F}(x,y)=                  
 \delta^{(4)}(x-y) \;.                                                         
\label{lightprop2}                                                              
\end{equation}
Our photon fields enter via
perturbative  QED, switched on by the replacement
$\partial_\mu \rightarrow D_\mu = \partial_\mu + i e A_\mu$ 
in (\ref{CoulDir}).
 It should be emphasized that $A_{\mu}(x)$ represents the radiation field and
does not include binding forces, which will in the next section 
be related to the external fields $V_{(i)}$.

Now, using relations (\ref{CoulDir}) and (\ref{lightprop}) we obtain
\beq
A_b  = -A_a^L + \Delta A_b \; , \qquad A_c  = -A_a^R + \Delta A_c  \;,
\eeq
resulting in a partial cancellation when the amplitudes are summed
\beq
A_a + A_b  + A_c =  \Delta A_b + \Delta A_c \; .
\eeq
This shows that the local off-diagonal fermion seagull 
transition of Fig. \ref{seagull} cancels, even if the external
fermions are off-shell.
The left-over  quantities $\Delta A_b$ and  $\Delta A_c$ involve 
the integrals over the Coulomb potential and represent the net
off-shell effect.

\begin{figure}
\centerline{\includegraphics[scale=0.7]{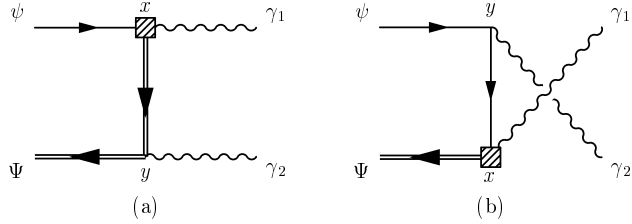}}
\caption{The shaded boxes indicate the combination of the
unrotated off-shell transition (proportional to $B_F$) and the on-shell magnetic
moment transition (proportional to $B_{\sigma}$), giving the effective vertex
in Eq. (\ref{shaded}).\label{MbMc}}
\end{figure}

There are also amplitudes $A_d$ and $A_e$ which are counterparts 
of $A_b$ and $A_c$
when ${\cal{L}}_F(1 \gamma)$ is replaced by ${\cal{L}}_{\sigma}$.
The total contribution from our flavour-changing Lagrangian 
(${\cal L}_F$ and ${\cal L}_\sigma$ parts) is then given by
\begin{eqnarray}
A_d + \Delta A_b & = &
\, \rmi    \int \!\! \int \rmd^4x\, \rmd^4y\,  \bar{\Psi}(y) 
\Big[-\rmi e A\slh_2(y)\Big] \hspace*{5em} \nonumber \\
& &\hspace*{5em}{} \times  \rmi S_F^{(\mu)}(y,x) \, Q(x) \, \psi(x) \; , 
\label{eq:lag1gDd}
\end{eqnarray}
represented by Fig. \ref{MbMc}a, and a similar one
\begin{eqnarray}
A_e + \Delta A_c  & = &
\,  \rmi   \int \!\! \int \rmd^4x\, \rmd^4y \, \bar{\Psi}(x) \, Q(x) \,    
 \rmi S_F^{(e)}(x,y)  \hspace*{3em} \nonumber \\
& & \hspace*{5em}{} \times  \Big[-\rmi e A\slh_2(y)\Big]\psi(y) \; ,
\label{eq:lag1gDe}
\end{eqnarray}
corresponding to Fig. \ref{MbMc}b. The operator $Q(x)$ in these expressions 
reads
\begin{eqnarray} 
\lefteqn{Q(x)  =   \big[ B_{\sigma}M + B_F V_{(\mu)}(x) \big] \,
  \sigma \cdot  F_1(x) L } \hspace*{5em}  \nonumber \\
& & {}+ \sigma \cdot  F_1(x) R \, \big[ B_{\sigma}m + B_F V_{(e)}(x) \big]\,.
\label{eq:Qexpr}
\end{eqnarray}
The result given by Eqs. (\ref{eq:lag1gDd})--(\ref{eq:Qexpr}) 
can also be understood in terms of the following
field redefinition.
Eq. (\ref{CoulDir}) can be obtained from the Lagrangian
\begin{equation}
{\cal{L}}_{D}(\Psi,\psi) = 
\bar{\Psi} \big[ \rmi D\slhh - V_{(\mu)} - M \big] \Psi +
\bar{\psi} \big[ \rmi D\slhh - V_{(e)} - m \big] \psi \; . 
\label{AppDir}
\end{equation}
Now, by defining new fields
\beq
\Psi ' = \Psi + B_F\,  \sigma \cdot  F \,  L  \psi \; , \qquad 
\psi ' = \psi + B_F ^* \,  \sigma  \cdot F  L \Psi \; ,
\label{repsi}
\eeq
we obtain
\beq
{\cal{L}}_{D}(\psi,\Psi) +  {\cal{L}}_{F}  \, = \,
{\cal{L}}_{D}(\psi ' , \Psi ') \, + \Delta {\cal{L}}_{B} \;,
\label{Trlag}
\eeq
which shows that ${\cal{L}}_{F}$ can be transformed away from 
the perturbative terms,
but a relic of it,
\beq
\Delta  {\cal{L}}_{B} \, = \, 
  B_F\, \bar{\Psi} \big[ V_{(\mu)} \sigma \cdot  F  L  \, + \,
  \sigma \cdot F R \,  V_{(e)} \big] \psi  \, + \textrm{h.c.} \; , 
\label{newlag}
\eeq
remains in the bound-state dynamics as a contact term. 
Thus, the off-shell effects
are non-zero for bound external fermions. Combining
$\Delta {\cal{L}}_{B}$ and ${\cal{L}}_{\sigma}$, we obtain
\beq
\Delta {\cal{L}}_{B} \, + \,{\cal{L}}_{\sigma} \; = \; 
\bar{\Psi} \, Q \, \psi \, + \textrm{h.c.} \; ,
\label{TotB}
\eeq
where $Q$ is given by (\ref{eq:Qexpr}). This shows how the
upper field redefinition rotates away the contact term shown in
Fig. \ref{seagull}, leaving us with the result
given by Eqs. (\ref{eq:lag1gDd})--(\ref{eq:Qexpr}).

\section{Off-shellness in the muonium annihilation amplitude}

The preceding section shows how far we can push the problem within
quantum field theory.
Up till now we have made no approximations except for standard 
perturbation theory.
Now we apply the obtained results to the double radiative
annihilation of muonium.
Naively, the product $\bar{\Psi} \, \psi$ corresponds to the bound state
of $\mu^+$ and $e^-$, which might be true only for the asymptotic free
fields. However, relativistic bound state physics is a difficult
 subject, which we circumvent by sticking to the two-step procedure
\cite{ItZ80} as explained in Section 1.
We perform the calculations in the muonium rest frame (CM frame of
 $\mu^+$ and $e^-$) where we put the external field(s) equal
to a mutual Coulomb field, $V_{(i)} \rightarrow \gamma_0 \, V_C$
(where $V_C = -e^2/4 \pi r$). In calculating the
$\mu^+ e^- \rightarrow \gamma \gamma$ amplitude in momentum space,
we take for $V_C$ the 
 average over  solutions in the
Coulomb potential,
 which is
$\langle V_C\rangle = - (m\alpha^2/2)$.
 In this way the 
muonium-decay invariant amplitude acquires the form which is
a straightforward generalization of the positronium-decay
invariant amplitude (\ref{positronium}) in momentum space.

The amplitudes $A_d + \Delta A_b$ from Eq. (\ref{eq:lag1gDd}), together
with $A_e + \Delta A_c$ from (\ref{eq:lag1gDe}), transformed to
the momentum space take the form
\begin{eqnarray}
{\cal M}&=&\frac{2 e B_{\sigma}}{m}\bar{v}_{\mu}(p_2)\Big\{
\frac{m}{M}k\slh_{2}\epsilon\slh_{2}^{*}{\rm P}-
{\rm P}\epsilon\slh_{2}^{*}k\slh_{2} + (1\leftrightarrow 2)
\Big\}
   u_{e}(p_1),
\label{5slashes}
\end{eqnarray}
where $v_{\mu}$ and $u_{e}$ are muon and electron spinors, and 
$\epsilon_{1,2}^{*}$ are photon polarization vectors.
The factor, incorporating the binding in the form of a four-vector
$U^{\alpha}=(\rho,\vec{0})$,
\begin{eqnarray}
{\rm P}\equiv(1-xU\slhh)k\slh_{1}\epsilon\slh_{1}^{*}L +
         xk\slh_{1}\epsilon\slh_{1}^{*}R(1-U\slhh) \;,
\end{eqnarray}
accounts for the aforementioned factorization of a binding and a decay,
and is represented by the shaded box of Fig. \ref{MbMc}:
\begin{equation}
 \Bigg[ M(1-x \rho \gamma^{0})\sigma \cdot F_{1}L + 
 m \,\sigma \cdot F_{1}R(1-\rho \gamma^{0}) \Bigg] \;.
\label{shaded}
\end{equation}
Here we introduced abbreviations for two small constant parameters,
\begin{equation}
x \equiv \frac{m}{M} \;, \qquad 
\rho \equiv - \, \frac{B_{F} \, \langle V_C \rangle}{m B_{\sigma}} \;,
\end{equation}
in terms of which the sought off-shell effect will be expressed.
Note that in the effective interaction (\ref{shaded}), the left-handed
part corresponding to $V_{(\mu)}$ has gotten an extra suppression
factor $x=m/M$ in front of the binding factor $\rho$, in agreement with
the expectation that the heavy particle ($\mu^+$) is approximately free,
and the light particle ($e^-$) is approximately the reduced particle,
in analogy with the H-atom.

The annihilation amplitude (\ref{5slashes}) can now be evaluated explicitly. 
The usual procedure of squaring the amplitude and using the Casimir trick
for converting spinors into Dirac matrices would give us expressions with
traces of up to twelve Dirac matrices, making the calculation unnecessary
extensive. It is much easier to proceed by going into the frame in which
the muonium is at rest and photons are emitted along the $z$-axis, i. e.
\begin{equation}
 k_1 = \left( \begin{array}{r}
       \omega \\ 0 \\ 0 \\ \omega
              \end{array} \right) \;, \quad
 k_2 = \left( \begin{array}{r}
       \omega \\ 0 \\ 0 \\ -\omega
              \end{array} \right) \;, \quad
 \epsilon_\pm = \frac{1}{\sqrt{2}} \left( \begin{array}{r}
       0 \\ 1 \\ \pm \rmi \\ 0
              \end{array} \right) \;,
\end{equation}
where $\omega=(m+M)/2\approx M/2$ is the photon energy.
In this frame $k\slh_i$ and $\epsilon^*\slhhh_j$ ($i,j=1,2$) formally 
anticommute
\begin{equation}
 k\slh_i \epsilon^*\slhhh_j = \omega(\gamma^0\pm\gamma^3)\frac{1}{\sqrt{2}}
 (\gamma^1\pm \rmi\gamma^2) = - \epsilon^*\slhhh_j k\slh_i  \;,
\end{equation}
so we can group them together and calculate
{\renewcommand{\arraystretch}{2}
 \setlength{\arraycolsep}{1.5ex}
\begin{equation}
 k\slh_1 k\slh_2 \epsilon^*\slhhh_1 \epsilon^*\slhhh_2 =\! - 2 \omega^2 \!
\left( \begin{array}{cc}
   \vec{\epsilon}_{2}^{*}\cdot\vec{\epsilon}_{1}^{*} -
(\vec{\epsilon}_{2}^{*} \times\vec{\epsilon}_{1}^{*}) \cdot\hat{\vec{k}}_{1}
\,\sigma^3   & 
\vec{\epsilon}_{2}^{*}\cdot\vec{\epsilon}_{1}^{*}\,\sigma^3 -
(\vec{\epsilon}_{2}^{*} \times\vec{\epsilon}_{1}^{*}) \cdot\hat{\vec{k}}_{1} \\
\vec{\epsilon}_{2}^{*}\cdot\vec{\epsilon}_{1}^{*}\,\sigma^3 -
(\vec{\epsilon}_{2}^{*} \times\vec{\epsilon}_{1}^{*}) \cdot\hat{\vec{k}}_{1} &
\vec{\epsilon}_{2}^{*}\cdot\vec{\epsilon}_{1}^{*} -
(\vec{\epsilon}_{2}^{*} \times\vec{\epsilon}_{1}^{*}) \cdot\hat{\vec{k}}_{1}
\,\sigma^3 \end{array} \right ) \,.
\end{equation}
}
It is now easy to multiply this by the appropriate chiral projectors $L$ and
$R$, $\rho\gamma^0$ terms, and $\bar{v}_{\mu}(p_2)$ and $u_{e}(p_1)$ spinors. 
Now, taking into account that muonium leading to the two-photon final state
is in the spin singlet state, we get the result
\begin{eqnarray}
{\cal M} &=& - 2 e B_{\sigma}  M^2 \sqrt{\frac{2M}{m}} \Big[
 (1-x^2+x\rho+x^2\rho)\vec{\epsilon}_{2}^{*}\cdot\vec{\epsilon}_{1}^{*} 
\nonumber \\
&& \hspace*{-1em}{}+\rmi(1+2x+x^2+x \rho-x^2\rho)(\vec{\epsilon}_{2}^{*}
  \times\vec{\epsilon}_{1}^{*})
\cdot\hat{\vec{k}}_{1} \Big] \;,
\label{Mamp}
\end{eqnarray}
In comparison to the expressions (\ref{positronium}) and
(\ref{parapositronium}) for parapositronium,  we notice that in addition to
$\vec{\epsilon}_{2}^{*}\times\vec{\epsilon}_{1}^{*}$ there appears also
$\vec{\epsilon}_{2}^{*}\cdot\vec{\epsilon}_{1}^{*}$, a CP-even two-photon
configuration.

The explicit expression for $\rho$ depends on some assumptions.
As explained previously, we use $\langle V_C \rangle = -m\alpha^2/2$ 
which gives
$\rho = \alpha^2/2$ for $B_\sigma = B_F = B$, which is a good 
approximation in the leptonic case.

\noindent
Eq.(\ref{width}) finally gives
\begin{equation}
\Gamma = \frac{2 \alpha  M^4}{m^2}|\psi(0)|^2 |B_{\sigma}|^{2}
  \left(1+ 2 x \rho \right) \;,
\label{totalG}
\end{equation}
where we have kept only the leading term in $\rho$ and $x$.
Since the wave function at the origin appears as a prefactor,
it is not necessary to know 
the precise value of $|\psi(0)|^2 \sim (m\alpha)^3/\pi$, in order to 
know the relative off-shell contribution.
Thus, for muonium, the sought off-shell contribution is only a tiny 
correction,
$2 x \rho = \alpha^2 m/M\simeq 2.6 \cdot 10^{-7}$, to the magnetic
moment dominated rate.

We may note in passing that we have checked our results also by the
direct calculation of the squared Feynman amplitude (\ref{5slashes})
on the computer using the \texttt{FeynCalc} Mathematica package for
algebraic manipulation of expressions involving Dirac matrices and
spinors \cite{MeBD91,Wo88}.
Here the explicit Lorentz covariance was preserved at all steps of the
calculation and the final result was in agreement with the one obtained by
hand calculation.

\section{Off-shellness in $\bar{B}^{0}_{s} \to \gamma \gamma$}

 In comparison to a tiny effect in the preceding section, we expect the
corresponding off-shellness in a strongly bound QCD system
to be significantly larger.
We also take into account the
$B_{F}/B_{\sigma}$ correction in (\ref{totalG}),
when considering the $\bar{B}^{0}_{s}\to\gamma\gamma$ decay.

The expressions
(\ref{eq:lagg}) to (\ref{eq:lag2g}) apply to the $b\rightarrow s \gamma \gamma$ 
induced $\bar{B}^{0}_{s}\rightarrow 2 \gamma$ decay amplitude by simple
replacements $\mu \rightarrow s$ and $e \rightarrow b$. Then one has
to scale the operators ${\cal{L}}_{F,\sigma}$ defined at the $M_{W}$ scale, 
down to the $B$-meson scale. 
The coefficients 
  $B_F$ of ${\cal{L}}_{F}$, and $B_{\sigma}$ of  ${\cal{L}}_{\sigma}$,
in Eqs. (\ref{eq:laggg}) and (\ref{eq:lagmag}),
 both being
equal to $B$ at the $W$ scale, may evolve differently down to the
 $\mu = m_b$ scale.
This difference between $B_{F}$ and $B_{\sigma}$ is due to different anomalous 
dimensions of the respective operators.
Within the SM one can write
\beq
B_{\sigma,F} = \frac{4 G_F}{\sqrt{2}} \, \lambda_{\rm CKM} \,
  \frac{e}{16 \pi^2} \, C_7^{\sigma,F} \; .
\label{defhat}
\eeq
The coefficient $C_7^\sigma$ has been studied by various authors
\cite{GrSW90,BuMMP94,GrOSN88,CeCRV90,CeCRV94}. 
The coefficient $C_7^F$ was considered in \cite{EeP94b}, where at
 the $b$-quark scale we obtained
\beq
\frac{C_7^{F}}{C_7^{\sigma}} \simeq 4/3   
\qquad  (\mu =m_{b}) \; .
\label{Bratio}
\eeq
Although the off-shell effect for $\bar{B} \rightarrow 2 \gamma$ 
is expected to be suppressed by the ratio (binding
energy)/$m_{b}$, it could still be numerically interesting.

The conventional procedure when evaluating the pseudoscalar meson
decay amplitudes is to express them in terms of the meson decay constants, 
by using the PCAC relations
\begin{eqnarray}
\langle 0 | \bar{s} \gamma_{\mu} \gamma_{5} b | \bar{B}^{0}_{s}(P) \rangle &=& 
- {\rmi f_{B} P_{\mu}} \;, \label{PCAC} \\
\langle 0 | \bar{s}  \gamma_{5} b | \bar{B}^{0}_{s}(P) \rangle  & = &
{\rmi f_{B} M_{B}} \;.
\label{Axcurrent}
\end{eqnarray}
These relations will be useful after reducing our 
general expression (\ref{5slashes})
containing the terms with products of up to five Dirac matrices. After
some calculation we arrive at the expression for the $\bar{B}_s$ meson decay
at rest, which is analogous to, and in fact confirms our previous
relation (\ref{Mamp}) obtained in a different way,
\begin{eqnarray}
{\cal M}^{B} &=& - \rmi \frac{e}{3} B_{\sigma} f_{B} M^2 
\frac{(1+x)^2}{x} \Big[
 (1-x^2 +x \tau +x^2 \tau)\,\vec{\epsilon}_{2}^{*}\cdot\vec{\epsilon}_{1}^{*} 
  + \nonumber \\
 & + & \; \rmi(1+2x +x^2 +x \tau -x^2 \tau)(\vec{\epsilon}_{2}^{*}
  \times\vec{\epsilon}_{1}^{*})
\cdot\hat{\vec{k}}_{1} \Big].
\label{Mampb}
\end{eqnarray}
Here, the parameter $\tau$ represents the off-shell effect in the
QCD problem at hand,
and will be more model dependent than its QED counterpart $\rho$.
With the amplitude (\ref{Mampb}), keeping only the leading terms in
$\tau$ and $x$,
we arrive at the total decay width
\begin{equation}
\Gamma = \frac{\alpha  M^5}{18 m^2} f_{B}^{2}
 |B_{\sigma}|^2 \left(1+ 2 x \tau \right) \;,
\end{equation}
where by switching off  $\tau$ we reproduce the result of Ref.
\cite{DeDL96}.

\subsection{Coulomb-type QCD model}

In order to estimate the value of the off-shell contribution $\tau$, in
this subsection we assume a QED-like QCD model with the Coulombic wave
function \cite{Go97,TrM88}
$\psi(r)\propto \exp(-m r \alpha_{\rm eff})$.
Thus we rely again on an exact solution cor\-res\-pon\-ding to effective 
potential
$V(r) = - 4\alpha_{\rm eff}/(3 r)$, with effective coupling 
$\alpha_{\rm eff}(r)\! = \! -(4\pi b_{0} \ln (r\Lambda_{\rm pot}))^{-1}$. 
%\\
Here $b_0=(1/8\pi^2)(11-(2/3)N_f)$.  The mass scale $\Lambda_{\rm pot}$
appropriate to the heavy-light quark $\bar{Q}q$ potential
 is related to the more familiar
QCD scale parameter, e.g. $\Lambda_{\rm pot}=2.23\,
  \Lambda_{\rm \overline{MS}}$
(for $N_f$=3). Within this model, we obtain
\begin{equation}
\tau =\frac{2}{3}\alpha_{\rm eff}^2 \frac{C_{7}^{F}}{C_{7}^{\sigma}}
\;.
\end{equation}
By matching the meson decay constant
$f_{B}$ and the wave function at the origin
\begin{equation}
N_{c}\frac{|\psi_{B}(0)|^2}{M}=\left(\frac{f_B}{2}\right)^2 \; ; \quad
|\psi_{B}(0)|^2=\frac{(m \alpha_{\rm eff})^3}{\pi} \;,
\end{equation}
we obtain the value for the strong interaction fine structure 
strength $\alpha_{\rm eff}$ $\approx$1.
Then, including (\ref{Bratio}) for the QCD case, the correction
factor
\begin{equation}
 x \tau \approx \, 0.1\,  \;,
\label{QCDoff}
\end{equation}
is much larger than $x\rho$ in the corresponding QED case.
Correspondingly, one expects even more significant off-shell effects
in light quark systems, in compliance with our previous
results \cite{EeP93,EeP94,KeKKP95}.

\subsection{A constituent  quark  calculation}

As an alternative to the Coulomb-type QCD model described above, now we
adopt a variant of the approach in Refs. \cite{EeP94b,VaE98}. One might
use the PCAC relations
(\ref{PCAC})--(\ref{Axcurrent}) together with a kinematical
assumption for the $\bar{s}$-quark momentum, similar to those in Refs.
\cite{DeDL96,HeK92}. We assume the bound $\bar{s}$ and
$b$ quarks in $\bar{B}^{0}_s$ to be on their respective \emph{effective}
mass-shells.
Note that even if one is using (\ref{PCAC}) and (\ref{Axcurrent}),
the amplitude will still explicitly depend on the $\bar{s}$-quark
momentum $p_{\bar{s}}$. This is put on the effective mass-shell by
using the relation $p_{\bar{s}}^{\mu}=-M_s (k_1 + k_2)^{\mu} / M_b $,
where $M_q = m_q + m_0$ (for $q=b,s$) are the effective (total) 
masses, $m_q$ are the current masses, and $m_0$ the constituent
mass of order a few hundred MeV.
The structure of the amplitude now comes out essentially as in
(\ref{Mampb}) with a relative off-shell contribution
\begin{equation}
x \, \tilde{\tau} \, = \, \frac{2 m_0}{m_b} \, \approx 0.1 \; ,
\label{Bsoff}
\end{equation}
of the same order as in (\ref{QCDoff}). However, unlike (\ref{Mampb}), the
off-shell effect is now only in the CP-odd term 
($\vec{\epsilon}^{\ast}_{1}\times\vec{\epsilon}^{\ast}_{2}$), the
square bracket in (\ref{Mampb}) being replaced by
\begin{equation}
\Big[\vec{\epsilon}_{2}^{*}\cdot\vec{\epsilon}_{1}^{*} 
  + 
 \rmi(1+2x+x \tilde{\tau})(\vec{\epsilon}_{2}^{*}
  \times\vec{\epsilon}_{1}^{*}) \Big] \;.
\end{equation}
This may
be different in other approaches \cite{EeP94}, showing the model dependence
of the off-shell effect. For instance, potential-QCD models in general, besides
a vector Co\-u\-lomb potential, also contain a scalar potential.

\subsection{A bound state quark  model}

\begin{figure}
\centerline{\includegraphics[scale=1.0]{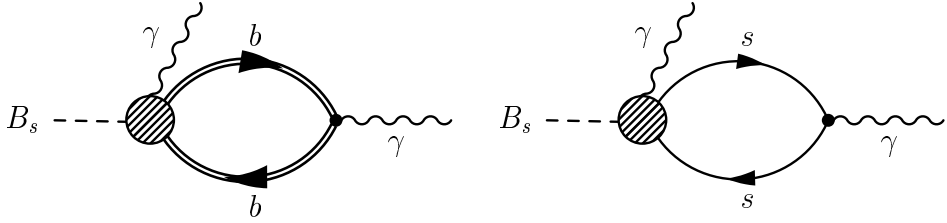}}
\caption{The two-photon transition amplitude from a contact term (\ref{newInt})
 left over after the field redefinitions.\label{contact}}
\end{figure}

In our previous accounts
\cite{EeP94b,VaE98} 
 we applied  a bound state model for $\bar{B}^{0}_{s} \rightarrow 2 \gamma$. 
Then the potentials $V_{i}$ in
 (\ref{CoulDir}) are  replaced by a quark-meson interaction 
Lagrangian
\beq
{\cal{L}}_{\Phi}(s,b) \, = 
\, G_B \bar{b} \, \gamma_5 \, s \, \Phi + 
  {\rm h.c.} \; ,
\label{qmes}
\eeq
where $\Phi$ is the B-meson field. In this case, the term ${\cal{L}}_{F}$ can be
 transformed away by means of the  field redefinitions:
\beq
 s' = s + B_F\,  \sigma \cdot  F \,  L  \, b \; , \qquad 
 b' = b + B_F ^* \,  \sigma  \cdot F  L \, s \; .
\label{rebs}
\eeq
However, its 
 effect  reappears in a new bound-state
 interaction $\Delta {\cal{L}}_{\Phi}$,
\beq
{\cal{L}}_{\Phi}(s,b)   + {\cal{L}}_{F} \, = \,
{\cal{L}}_{\Phi}(s',b') \, + \Delta {\cal{L}}_{\Phi} \;,
\label{DeltaL}
\eeq
where, after using $R \gamma_5 = R$ and $L \gamma_5 = -L$,
\beq
\Delta  {\cal{L}}_{\Phi} \, = \, 
  B_F \, G_B \,  \big[ \bar{b}' \sigma \cdot  F  L  \, b' \, - \,
  \bar{s}'  \, \sigma \cdot F R \, s'  \big] \Phi   \, + {\rm h.c.} \;.
\label{newInt}
\eeq
The two terms in this equation correspond to two contact amplitudes
displayed in Fig. \ref{contact}.
Also in this case, net off-shell effects are found \cite{EeP94b,VaE98}.
Further calculations of  $B \rightarrow 2\gamma$ within bound state
models of the type in (\ref{qmes}) will be presented elsewhere.

 Note that in bound-state models  based on heavy-quark
effective theory the expression (\ref{qmes}) is slightly modified such that
the $b$ quark field will be replaced by the product of the
reduced heavy-quark field and its projector 
$P_+(v) = (1+\gamma \cdot v)/2$,
where $v$ is the velocity of the heavy quark 
 \cite{BaH94,EbFFR95,DeBG98,Po00}.

\subsection{Link to $\bar{B}_d \rightarrow K^* \gamma$}

Although we focused till now only to the two-photon processes, the
interaction in (\ref{eq:laggg}) contributes to the one-photon
couplings as well.

Actually, we observe that off-diagonal one-photon couplings contained
in the Lagrangian
given by (\ref{eq:Qexpr}) and (\ref{TotB}) can be used  
to calculate the amplitude for muonic hydrogen
decaying to a photon and ordinary hydrogen, that is, the process 
$\mu^- \rightarrow e^- + \gamma$ for both leptons bound to
a proton. This is a leptonic version of the celebrated $B$-meson decay
$\bar{B}_d \rightarrow K^* \gamma $.
 
As a toy model, one might consider a process
 ``$\mu$'' $\rightarrow$ ``$e$'' $\gamma$ in an external Coulomb field,
with ``$\mu$'' and ``$e$'' rather close in mass such that the
non-relativistic descriptions of the ``leptons'' might be used.
The effective ``$\mu$'' $\rightarrow$ ``$e$'' $\gamma$ interaction is
given in (\ref{TotB}). If we assume that $(M-m)$ is of order $\alpha \, m$,
we obtain  off-shell effects of order $\alpha^2$
due to ${\cal{L}}_F$, relative to the standard magnetic moment
term ${\cal{L}}_\sigma$.
Bigger mass 
differences give bigger effects, until the 
non-relativistic approximation breaks down.

Returning to the 
off-shell bound-state effects in the important $B_{d}\to K^{*}\gamma$
decay,
% we notice that they have been considered in a preprint version
%of Ref. \cite{CaM95}.
they can be addressed in the  framework of
models \cite{BaH94,EbFFR95,DeBG98,Po00,EeFZ01,HiE01} combining 
heavy quark effective theories with the ideas of Nambu-Jona-Lasinio models
and chiral quark models.

\begin{figure}
\centerline{\includegraphics[scale=1.0]{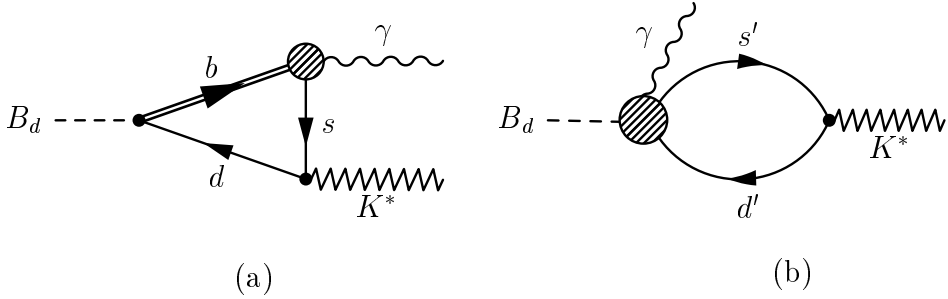}}
\caption{Diagrams for $\bar{B_d} \rightarrow \gamma K^*$: (a)
The magnetic moment transition amplitude. (b) The contact term
(\ref{newcontact}) left over after field redefinitions.\label{bkg}}
\end{figure}

The ordinary, on-shell transition magnetic moment ${\cal L}_{\sigma}$-induced
amplitude for the $\bar{B}_d \to K^* \gamma$ is shown in
Fig. \ref{bkg}a.

Now, transforming away the term ${\cal L}_F$ by the field redefinitions
produces the new contact term
\begin{equation}
 \Delta {\cal L}_{\Phi}' = -B_F \,  G_B \, \bar{s}' 
\sigma\cdot F \,  R \, P_{+}(v) \gamma_5 \, d' \, \Phi 
\label{newcontact}
\end{equation}
giving the amplitude displayed in Fig. \ref{bkg}b.

The ratio of the off-shell and the on-shell amplitudes 
in the soft $K^*$ limit can now be calculated  to be
\begin{equation}
  \frac{A(\bar{B}_d\to K^* \gamma)_{{\rm off-shell}}}
       {A(\bar{B}_d\to K^* \gamma)_{{\rm on-shell}}}
\approx \frac{B_F}{B_\sigma} \frac{m_0}{2m_Q} \;,
\label{rez}
\end{equation}
where $m_0$ is the constituent mass of order of 200--300 MeV, and
$m_Q$ is the heavy ($b$ quark) mass.
Going away from the soft $K^*$ limit, the amplitudes will change, but
the result (\ref{rez}) will persist in the leading order.

A refined calculation would be desirable in view of the importance
of this result. An earlier attempt (the preprint version of
Ref. \cite{CaM95}) reported on large off-shell effect in the amplitude
which happened to be reduced below 10 \% effect in the rate.

\section{Discussion and conclusion}

The present ``atomic'' approach enables us to see in a new light the
off-shell effects 
studied first for the $K_L\rightarrow  \gamma\gamma$ amplitude in the chiral 
quark model \cite{EeP93,EeP94}, and subsequently in the bound-state model 
\cite{KeKKP95}. The observation that off-shell effects can be clearly isolated 
from the rest in the heavy-light quark atoms \cite{EeP94b} was still plagued 
by the uncertainty in the QCD binding calculation \cite{VaE98}. 
Here, in the Coulomb-type QCD model we are able to subsume the effect into 
an universal binding factor, in the same way as for the two-photon decay of 
muonium in the exactly solvable QED framework.
It is a quite significant 10 percent effect in the $\bar{B}^{0}_{s}
\rightarrow \gamma \gamma$ case, whereas in the two-photon decay of
muonium it is very small (of order 10$^{-7}$), but clearly identifiable.

As a byproduct we obtain here also the on-shell amplitude
already considered in the literature.
There is an extensive list of calculations \cite{DeDL96,HeK92,ReRS97}
pertinent to the
short distance electroweak loop contributions to $b\to s\gamma\gamma$ which
trigger $\bar{B}_s \to \gamma\gamma$. Comparing our results to the expression
(22) of \cite{DeDL96} we can express $C_{7}^{\sigma}$ in our (\ref{defhat})
in terms of their coefficient C
\begin{equation}
   C^{\sigma}_7 = \frac{1}{4\sqrt{6}}\left( C + \frac{23}{3} \right) \;,
\end{equation}
or, numerically, $C^{\sigma}_7=0.4$ at the $B$ meson scale.
Still, there is another class of contributions, belonging to the LD
regime. For example \cite{Hi97,Hi98} present magnetic moment,
 $O_7$-type LD effects in
$\bar{B}_s \to \gamma\gamma$ decays in the vector meson dominance approach,
whereas the other authors \cite{ChE98,LiZZ99,Mi99}, though with
controversial results, estimate the contribution of the charmed-meson
intermediate states. These seem to be a natural representation for
our short-distance loops when the loop momenta are below the
$b$ quark mass scale.

Our message is that such small SM effects might
obscure possible new physics (BSM) signals that are of a comparable size. 
Without pretending on completeness we give some examples that
the off-shell effects considered in the present paper might
hinder possible BSM discovery.

Let us start with the famous magnetism of the muon, an ideal of
the precision measurement. At some level the binding effects might
become relevant.
Such
$(g_{\textrm{\scriptsize Bound}}-2)$ effect due to the diagonal
one-photon coupling would correspond to the 
$(g_{\textrm{\scriptsize Bound}}-2)$
calculated already for a bound electron \cite{Cz00}.
This effect might be interesting in light of a deviation
from the Standard Model expectation of the
order of $10^{-9}$ recently measured for $(g-2)$ of the positively charged
muon \cite{Br01}. Actually, this measurement triggered various speculations
ascribing this discrepancy to the various
BSM effects, the lepton compositeness \cite{CaFH01}
being one possibility.
However, there are more direct ways to set a bound on the compositeness
scale from the flavour-conserving processes.
For example, there are flavour-diagonal $e^+ e^- \gamma\gamma$ contact terms
\cite{Ba00}
\begin{equation}
{\mathcal L}_{\textrm{\scriptsize contact}}=\rmi\overline{\psi_{e}}
\gamma_{\mu}(D_{\nu}\psi_{e})
\left(\frac{\sqrt{4\pi}}{\Lambda^{2}_{6}}F^{\mu\nu}+
\frac{\sqrt{4\pi}}{\tilde{\Lambda}^{2}_{6}}\tilde{F}^{\mu\nu}\right)
\label{eq.2}
\end{equation}
which would lead to a ($1+\delta_{\textrm{\scriptsize DEV}}$) correction
factor to the photon angular distribution $d\sigma/ d\Omega$ in
$e^+ e^-$ collisions.
From 
\begin{equation}
\delta_{\textrm{\scriptsize DEV}}=
s^{2}/(2\alpha)(1/\Lambda^{4}_{6} + 1/\tilde{\Lambda}^{4}_{6})
(1-\cos^{2}\theta)
\end{equation}
LEP200 sets a bound $\Lambda>1687$ GeV 
(for $\Lambda_6=\tilde{\Lambda}_{6}=\Lambda$) at 95 \% CL.
Thus, eventual non-standard BSM physics contribution at LEP energies are
highly suppressed.

More promising route to reveal BSM contributions could be provided by
flavour non-diagonal transitions. Recent evidence for neutrino oscillations
has renewed interest in charged LFV searches. Among variety of probes
reviewed in \cite{KuO99} $\mu\to e \gamma$ and $\mu - e$ conversion 
(invoking new high energy scale $M_{12}$) seem
to be the most promising.
Since the effects of new physics are expected to enter at one-loop level,
these transitions may be parameterized by
\begin{equation}
{\cal L}_{12} = e \frac{g^2}{16 \pi^2} \frac{m_{\mu}}{M_{12}^2}
\; \bar{\mu} \sigma^{\alpha \beta} e \; F_{\alpha \beta} \;,
\label{specific}
\end{equation}
in order to estimate the sensitivity of the current experimental facilities
\cite{Feng00}.

Although the BSM effects might be more pronounced for the flavour-changing
quark transitions, their discovery might be hindered by the
relatively more pronounced bound-state effects treated in the
present paper.
 The off-shell contribution may affect the discovery
potential in the radiative $B$ meson decays (for example, 
\cite{Gr98,Mi00,AhR01}).  
In particular, our result (\ref{rez}) indicates hindrance of the
BSM discovery potential in the otherwise promising 
$\bar{B}_d \to K^* \gamma$ decay.

\end{document}